\documentclass[sigconf]{acmart}
\usepackage{tikz}
\usetikzlibrary{calc, shapes, positioning, shadows, trees, backgrounds, matrix, chains, arrows, decorations.pathmorphing, decorations.pathreplacing, fit, bayesnet, arrows.meta}
\usepackage{blkarray}
\usepackage{array}
\usepackage{booktabs}
\usepackage{adjustbox}
\usepackage{csquotes}
\usepackage{array}
\usepackage{balance}

\AtBeginDocument{%
  \providecommand\BibTeX{{%
    \normalfont B\kern-0.5em{\scshape i\kern-0.25em b}\kern-0.8em\TeX}}}

\copyrightyear{2021}
\acmYear{2021}
\setcopyright{licensedothergov}\acmConference[SIGIR '21]{Proceedings of the 44th International ACM SIGIR Conference on Research and Development in Information Retrieval}{July 11--15, 2021}{Virtual Event, Canada}
\acmBooktitle{Proceedings of the 44th International ACM SIGIR Conference on Research and Development in Information Retrieval (SIGIR '21), July 11--15, 2021, Virtual Event, Canada}
\acmPrice{15.00}
\acmDOI{10.1145/3404835.3462994}
\acmISBN{978-1-4503-8037-9/21/07}

\begin{document}
\fancyhead{}
\renewcommand{\tabcolsep}{1pt}

\title{Predicting Links on Wikipedia with Anchor Text Information}

\author{Robin Brochier}
\email{robin.brochier@lis-lab.fr}
\orcid{0000-0002-6188-6509}
\affiliation{%
  \institution{Aix Marseille Univ, Universit\'e de Toulon, CNRS, LIS, Marseille, France}
  \streetaddress{Campus universitaire de Luminy,\\ Bat. TPR2, 5\`eme \'etage, Bloc 1,\\ 163 avenue de Luminy,\\ 13288 Marseille cedex 09}
  \country{France}
}

\author{Fr\'ed\'eric B\'echet}
\email{frederic.bechet@lis-lab.fr}
\orcid{0000-0002-2650-750X}
\affiliation{%
  \institution{Aix Marseille Univ, Universit\'e de Toulon, CNRS, LIS, Marseille, France}
  \streetaddress{Campus universitaire de Luminy,\\ Bat. TPR2, 5\`eme \'etage, Bloc 1,\\ 163 avenue de Luminy,\\ 13288 Marseille cedex 09}
  \country{France}
}

\renewcommand{\shortauthors}{Brochier et al.}

\begin{abstract}
  Wikipedia, the largest open-collaborative online encyclopedia, is a corpus of documents bound together by internal hyperlinks. These links form the building blocks of a large network whose structure contains important information on the concepts covered in this encyclopedia. The presence of a link between two articles, materialised by an anchor text in the source page pointing to the target page, can increase readers' understanding of a topic. However, the process of linking follows specific editorial rules to avoid both under-linking and over-linking. In this paper, we study the transductive and the inductive tasks of link prediction on several subsets of the English Wikipedia and identify some key challenges behind automatic linking based on anchor text information. We propose an appropriate evaluation sampling methodology and compare several algorithms. Moreover, we propose baseline models that provide a good estimation of the overall difficulty of the tasks.     
\end{abstract}


\maketitle

\section{Introduction}

Hyperlinks are the backbone of the Internet. Placed within the hypertext of web pages, they allow users to explore the World Wide Web by clicking on \textit{anchor texts} redirecting them to new contents of interest. Moreover, the global link structure of the Internet can be seen as a network carrying rich features for information retrieval systems such as web browsers \cite{page1999pagerank, kleinberg1999authoritative}. On Wikipedia, collaborative editors are asked to follow specific guidelines \footnote{\url{https://en.wikipedia.org/wiki/Wikipedia:Manual_of_Style/Linking}} to ensure the quality of internal links (or wikilinks). Indeed, these links should highlight the concepts presented in an article while not overwhelming the readers with unnecessary connections.

\begin{figure}[htb]
    \begin{quote}
        \small
        Abraham Lincoln (February 12, 1809 - April 15, 1865) was an \textbf{\{American|Americans\}} statesman and lawyer who served as the 16th \textbf{[president of the United States|President of the United States]} from 1861 until his assassination in 1865. Lincoln led the nation through the \textbf{[American Civil War|American Civil War]}, the country's greatest moral, constitutional, and \textbf{\{political|Politics\}} crisis. He succeeded in preserving the Union, abolishing \textbf{[slavery|Slavery in the United States]}, bolstering the \textbf{[federal government|Federal government of the United States]}, and modernizing the U.S. economy.
    \end{quote}
\caption{Example of an abstract taken from Wikipedia. Wikilinks are represented with the schema \textquote{[\textit{anchor text}|\textit{article title}]}. We added nonexistent links, between curly brackets \textquote{\{|\}}, whose anchor texts were found in other articles using a string-matching heuristic.}
\label{fig:wikilinks}
\end{figure}

Previous works have proposed algorithms to automatically identify new links in document networks: a link exists between two documents if their contents are semantically related. In this paper, we examine this task in the specific case of Wikipedia internal hyperlinks. 
In our case, a document $A$ is linked to document $B$ if there is a sequence of words (an anchor text) in $A$ which refers directly to document $B$.
Therefore, given a mapping from anchor texts to documents, the task of searching for new candidates can be tackled with a simple string-matching technique. The difficulty of the problem becomes the relevance of potential links given the topic addressed by the article i.e. if a potential anchor text is representative enough of the context of an article to be made an hyperlink.
Many sequences of words can be hyperlink candidates, e.g. when they match with titles of Wikipedia pages, however only a few of them will be actual links in the original document.

This work is a first step towards automatically predicting relevant hyperlinks for new documents with respect to a network of Wikipedia articles.
Our contributions are the following: \begin{itemize}
    \item we create and make publicly available \footnote{\url{https://github.com/brochier/wikipedia_hyperlink_networks}} several exports of Wikipedia centered around different topics in which the text content, the hyperlink network and the associated anchor texts of the articles are provided;
    \item we propose two evaluation protocols for a transductive and an inductive prediction tasks. In the former case, only a few links are missing from every articles while in the latter case, we aim at predicting all the links of previously unseen articles. Furthermore, we provide a strong evaluation sampling method that relies on false positive predictions of a simple string-matching algorithm; 
    \item we present experimental results performed with text-based, graph-based and hybrid methods for this task. 
\end{itemize}

\section{Related Works} \label{sec:rw}

Link prediction \cite{liben2007link, lu2011link} is a largely studied problem across various disciplines. In this paper, we focus on predicting links in a document network, i.e. when the nodes are associated with textual features. In this section, we first cover recent algorithms applied to the link prediction problem and then relate important works on Wikipedia's hyperlinks. 

\subsection{Link Prediction in Document Networks}

Early works on link prediction includes similarity-based methods \cite{adamic2003friends}, maximum likelihood models \cite{white1976social, clauset2008hierarchical} and probabilistic models \cite{friedman1999learning, heckerman2007probabilistic}. More recently, network embedding (NE) methods have achieved better performance and scalability in several application domains. DeepWalk (DW) \cite{perozzi2014deepwalk} and node2vec \cite{grover2016node2vec} are the most well-known NE algorithms. They train dense embedding vectors by predicting nodes co-occurrences through random walks by adapting the Skip-Gram model \cite{mikolov2013distributed} initially designed for word embedding. 

However, in document networks, these previous models do not benefit from the information contained in the text content associated with the nodes. To address this, several methods \cite{brochier2019global, gourru2020document} propose to combine text and graph features to produce efficient embeddings. As such, Text-Associated DeepWalk (TADW) \cite{yang2015network} extends DeepWalk to deal with textual attributes. They prove, following the work in \cite{levy2014neural}, that Skip-Gram with hierarchical softmax can be equivalently formulated as a matrix factorization problem. TADW then consists in constraining the factorization problem with a pre-computed representation of the documents by using Latent Semantic Analysis (LSA) \cite{deerwester1990ilsa}. Graph2Gauss (G2G) \cite{bojchevski2018deep} is an approach that embeds each node as a Gaussian distribution instead of a vector. The algorithm is trained by passing node attributes (document-term frequencies) through a non-linear transformation via a deep neural network (encoder). 
Inductive Document Network Embedding (IDNE) \cite{brochier2020inductive} is a method that produces, via a topic attention mechanism, representations of documents that reflect their connections in a network. In this direction, some approaches \cite{tu2017cane, brochier2019link} specifically model the textual features underlying each link in the network. 
Finally, Graph Neural Networks (GNN) \cite{scarselli2008graph, velivckovic2017graph, hamilton2017inductive} have recently shown great performances in link prediction problems. However, it is often non-trivial to infer new representations from the features only, since their aggregators rely on the neighborhood of a node. In this paper, we focus on methods that can easily induce predictions for new documents with no known link to the original network.   

Document network embedding methods have been applied to a wide variety of networks such as citation networks \cite{mccallum2000automating, gilesautomatic}, question \& answer websites \cite{brochier2020inductive} and newspaper networks \cite{gourru2020document}. However, Wikipedia has particular characteristics, the main one being the fact that links are materialised in the documents by anchor texts. Moreover, the existence of a link is the result of an editorial choice meant to improve the reader's ability to explore a particular topic. In the following section, we cover some of the works that studied these characteristics.  

\subsection{Wikipedia Hyperlink Network}

Improving link coverage \footnote{\url{https://meta.wikimedia.org/wiki/Research:Improving_link_coverage}} is an important challenge for any website, both to improve the navigation within that website and to enhance its referencing by external browsers. A direction for automatically finding useful hyperlinks relies on the user activity logs \cite{west2015mining, paranjape2016improving, gundala2018readers}. Given clickthrough rate of the hyperlinks on Wikipedia, it is possible to predict usefulness probabilities for candidate links. These candidates can be selected based on navigation paths, i.e. when users often navigate from a source page to a target page without using a direct hyperlink. Our approach differs since we seek to train a model to identify textual patterns that should produce a link where user logs are helpful to (1) identify candidates and (2) quantify the relevance of a link. Learning from text feature has the potential to generalize better particularly for real time prediction when a user publishes a new article. In this direction, \cite{wu2007autonomously} study the efficiency of a simple \textit{proper noun} based matching system, similar to those we use as baseline models in our experiments. The idea is to extract proper nouns from the article content and to match these with existing article titles. As we will show in Section \ref{sec:xp}, this tends to achieve low precision scores since it produces too many false positive links. Finally, it is well-established that Wikipedia gathers very different communities with different linking behaviors \cite{ruprechter2020relating}. To take this into consideration, we evaluate the algorithms on several subgraphs centered around different topics of the full Wikipedia network.             

\section{Predicting anchor text hyperlinks} \label{sec:ev}

The task of predicting if a sequence of words in a document is an anchor text linking to another document can be seen as a link prediction task in a network of documents.
To perform this task, as presented in Section~\ref{sec:rw}, classifiers are trained to predict a binary decision \textit{link/no link} with textual and graph information.
One common issue during this training phase is the collect of negative examples.

Sampling on all nonexistent links can be computationally infeasible as the densities of the networks are small (below 1\%, see Table \ref{tab:datasets}) and the number of potential documents pairs is the square of the number of documents. To reduce the amount of testing pairs, a usual methodology (used in \cite{grover2016node2vec}) consists in randomly sampling as many negative pairs as positive ones. This tends to produce extremely high and thus not meaningful scores as the link probability between two nodes in social networks is often highly correlated with the average path length between them.

In this paper we are interested in a specific kind of document linking task as all links must start from some anchor text pointing to another document in the network.
This is a strong constraint that can be used to select relevant negative samples based on a simple string-matching procedure, that is, we collect every anchor text responsible for a link in the network, and for each of them, we identify all the articles that contain the same sequence of words. By doing so, we generate negative samples that are hard to distinguish with positive ones in terms of text features.

One of the goals of this paper is to compare generic document linking methods to specific ones taking into account the specificities of anchor text hyperlink prediction. First, we present the strong baselines we used to perform document linking and then we introduce some novel simple models based on heuristics specific to the task targeted in this paper.   

\subsection{Document linking with textual and graph information}

We compared several algorithms to perform document linking prediction. The following methods rely either or both on the textual and graph information contained in the document network. Note that for all methods, we construct representations of dimension $d=512$:
\begin{itemize}
    \item LSA: we use an SVD decomposition of TF-IDF vectors (term frequency inverse document frequency \cite{jones1972statistical});
    \item DeepWalk (DW): we use skip-gram with $10$ negative samples (following \cite{grover2016node2vec}) and we apply a window of size $10$ after performing $80$ random walks per node of lengths $40$. Since DW makes no use of the text features, it cannot be used in the inductive case;  
    \item IDNE: we run all experiments with $n_t=32$ topic vectors performing $1000$ iterations with mini-batches of $64$ balanced positive and negative samples;
    \item Graph2gauss (G2G): we use term frequency representations of the documents as input;
    \item TADW: we follow the guidelines of the original paper by using 20 iterations and a penalty term $\lambda=0.2$. For induction, we follow the strategy presented in \cite{brochier2020inductive}.
\end{itemize}

\subsection{Document linking in the context of anchor text hyperlink prediction}

In addition to these methods, we developed two heuristic techniques based on anchor texts (AT), named AT (title) and AT (anchor), that rely on a string-matching procedure. Both methods are given a mapping from strings to Wikipedia articles. They only differ in the way this mapping is constructed:
\begin{itemize}
    \item AT (title): maps any article's title with its article. Moreover, any title redirecting to this article is also considered (e.g. both \textquote{United Kingdom} and \textquote{UK} will map the article \textit{United Kingdom}).  
    \item AT (anchor): maps any anchor text encountered in Wikipedia to the targeted article. This allows us to match all the hyperlinks of the datasets, ensuring the highest recall possible. However, this model achieves low precision as it tends to over-link.   
\end{itemize}
Given their respective mappings, these two algorithms predict the existence of a link between two articles if and only if the source page contains at least one of the strings mapping to the target page. As such, these methods output exact predictions (true of false). 

Finally, given that AT (anchor) achieves perfect recall, we propose a simple model, ATILP (Anchor Text Informed Link Prediction) that focuses on reducing the number of false positives retrieved by the former. This model selects all candidate links identified by AT (anchor), and extracts a representation of their anchor texts using LSA. Then, three scores are computed given the LSA vectors of the anchor texts $x_{at}$, the source documents $x_{ds}$ and the target documents $x_{dt}$:
\begin{itemize}
    \item $s_1$: is the cosine similarity between $x_{at}$ and $x_{ds}$, representing how the anchor text is similar to the source document;
    \item $s_2$: is the cosine similarity between $x_{at}$ and $x_{dt}$, representing how the anchor text is similar to the target document;
    \item $s_3$: is the cosine similarity between $x_{dt}$ and $x_{ds}$, representing how the two documents are similar. Note that this score is directly used a prediction by the LSA model.
\end{itemize}

We then train a least squares linear regression without neither normalization nor regularization on the previous scores, $(s_1, s_2, s_3)$, to predict the probability of a link between a pair of documents. The training set is built by randomly sampling $1000$ existing and $1000$ nonexistent links selected with AT (anchor). The motivation behind this simple model is that (1) $s_1$ should capture if an anchor text represents the concepts of the source document, (2) $s_2$ should represent how much the anchor text specifically describes the target document and (3) $s_3$ should indicate if the two documents are semantically related.    

\section{Evaluation setup}

The evaluation of link prediction methods \cite{yang2015evaluating} needs careful design. Our objective is to quantify the ability of an algorithm to identify relevant links in Wikipedia articles. Given a network of documents, an algorithm is trained based on text/graph features and is then asked to produce the probabilities of links given pairs of documents. We consider two cases for which we provide experiment results:
\begin{itemize}
    \item the transductive case: the algorithms are trained on a network and try to predict missing links within that network (increasing the number of links). To simulate this, we remove 10\% of the links from the original network during training and evaluate the algorithms on the hidden links. Note that in this case, an algorithm can leverage the existing links connecting a document for its predictions.
    \item the inductive case: the algorithms are trained on a network and try to predict links for new documents outside of the network (increasing the number of documents). To simulate this, we remove 10\% of the documents from the original network during training and evaluate the algorithms on the hidden documents. Note that in this case, an algorithm can only rely on the text features of a document to predict its links.  
\end{itemize}

In both cases, once the algorithms have produced link probabilities ($p \in [0,1]$) or absolute predictions ($p \in \{0,1\}$), we compute the area under the precision-recall curve (AUC) given the true labels ($1$ if a link exists, $0$ otherwise). Moreover, we report the precision (P) and the recall (R) of the predictions. When an algorithm outputs probabilities, we threshold them given the true number of positive samples in the test set, enforcing equal values of precision and recall. 


Evaluating both the transductive and inductive cases allows us to identify how well an algorithm can generalize from text features. 
However, we should only compare the rankings of the algorithms and not directly their scores since the ratio of negative to positive samples is higher in the transductive case than in the inductive one. This is due to the fact that we test only 10\% of the existing links in the first case and test all existing links of the hidden documents in the second case. This has the effect of producing lower scores for the transductive task, even if it is an easier task to solve. 

For all experiments, we run $5$ times the evaluations with the same splits for all algorithms and report the AUC, the precision and the recall in Section \ref{sec:xp}. For each dataset, we additionally report the scores obtained by a random algorithms for comparison. 

\subsection{Datasets}

To build the datasets, we download an XML dump of Wikipedia \footnote{\url{https://dumps.wikimedia.org/}} from January 2021 and we extract the full unweighted directed network of wikilinks with the abstract of each article. Moreover, we extract the anchor text of each link in these abstracts. Then, we construct subgraphs centered around $4$ articles dealing with a variety of topics, namely politics (Joe Biden), science (Science), literature (The Little Prince) and sport (Cristiano Ronaldo). The extraction is performed by computing the personalized PageRank scores of these articles and by then selecting the $1000$ highest ranked articles. The main properties of the resulting networks are reported in Table \ref{tab:datasets}.      

\begin{table}
\center
\small
\caption{ Datasets properties: numbers of documents $n_V$, of links $n_E$, and of words in the vocabulary $n_W$, average (and standard deviation) of document lengths $\ell_D$ and of the numbers of positive and negative samples per document $n_+$ and $n_-$ used for the evaluations.}
\label{tab:datasets}
\begin{adjustbox}{width=\columnwidth}
\begin{tabular}{r | c  c  c  c  c  c} 
               \textbf{  }             & \textbf{ $n_V$ } & \textbf{ $n_E$ }  & \textbf{ $n_W$ } & \textbf{ $\ell_D$ } & \textbf{ $n_+$ } & \textbf{ $n_-$ } \\ \hline
Joe Biden &    1,000     & 7,817 (0.78\%) &    18,078    &      306 (215)       &  12.18 (10.39)   &  30.40 (20.37)   \\
Science &    1,000     & 8,021 (0.80\%) &    19,645    &      284 (194)       &   11.90 (9.76)   &  32.38 (18.66)   \\ 
The Little Prince &    1,000     & 6,022 (0.60\%) &    22,051    &      283 (222)       &   8.77 (8.60)    &  21.61 (15.51)   \\ 
Cristiano Ronaldo &    1,000     & 7,048 (0.70\%) &    16,955    &      264 (297)       &  10.53 (10.02)   &  31.88 (22.02)   \\ 
\end{tabular}\end{adjustbox}
\end{table}

\section{Experiment Results} \label{sec:xp}

In Table \ref{tab:results}, we report the results of the experiments. As expected, AT (anchor) has a perfect recall but has a low precision. AT (title) shows that (1) not all anchor texts are as trivial as article's titles since the model achieves only 50\% recall and (2) many sequences of words corresponding to titles are not worth being made a link in the documents (low precision). 

Graph-based (DW) and text-based (LSA) models perform similarly across the datasets, with a small advantage for LSA. Hybrid methods (G2G, TADW and IDNE) do not bring any improvement. Our interpretation is that these methods are good for capturing general concepts, but they do not capture fine grained discriminative features capable of distinguishing two conceptually related pages from two actual linked pages. For Wikipedia, this distinction is hard to learn because it is mainly due to the editorial choice made by the collaborative users.  

Finally, ATILP achieves slightly better scores than the best models. The learned coefficients $c$ of the linear regression associated with the scores  $s_1, s_2$ and $s_3$ have average values across all runs: $c_{s_1} = 0.10 (0.09) $, $c_{s_2} = 0.36 (0.04)$ and $c_{s_3} = 1.06 (0.10)$. Since no normalization is applied beforehand, these coefficients show how much the scores weight in the estimation of the probability of a link. Without surprise, the semantic similarity $s_3$ between the two documents accounts for the majority of the prediction. However, the similarity $s_2$ between the anchor text and the target document seems to help the model to improve the predictions. We hypothesize that the model identifies anchor texts representing too general concepts such as \textquote{American} in Figure \ref{fig:wikilinks}, thus decreasing the predicted probability of a link when the similarity $s_2$ is low given a high $s_3$.  


\begin{table}
\center
\caption{ Experiment results. The AUC, precision, recall and their standard deviations (in parenthesis) for each dataset, task and method are reported.}
\label{tab:results}

\begin{adjustbox}{width=\columnwidth}
\begin{tabular}{ >{\centering\arraybackslash}p{2cm} | c  c  c | c  c  c } 
\textbf{ Joe Biden } & \multicolumn{3}{c|}{Inductive}  & \multicolumn{3}{c}{Transductive} \\
       &    \textbf{ AUC }    & \textbf{ P } &   \textbf{ R }   &    \textbf{ AUC }    & \textbf{ P } &   \textbf{ R }   \\ \hline
    random &       21.17 ( 1.03)      &       20.70 ( 1.33)      &       20.70 ( 1.33)       &        2.84 ( 0.22)      &        3.29 ( 0.33)      &        3.29 ( 0.33)       \\  \hline
    AT (title) &       34.08 ( 0.67)      &       47.28 ( 0.64)      &       49.53 ( 2.06)       &        5.30 ( 0.21)      &        8.01 ( 0.30)      &       50.46 ( 0.08)       \\ 
    AT (anchor) &       21.18 ( 0.81)      &       21.18 ( 0.81)      & \textbf{ 100.00 ( 0.00) } &        2.54 ( 0.10)      &        2.54 ( 0.10)      & \textbf{ 100.00 ( 0.00) } \\  \hline
    LSA &     53.25 ( 2.34)      &       51.69 ( 1.73)      &       51.71 ( 1.72)       &       13.29 ( 1.58)      &       20.20 ( 2.20)      &       20.20 ( 2.20)       \\  \hline
    DW &            -            &            -            &            -             &       12.34 ( 0.72)      &       19.07 ( 1.84)      &       19.10 ( 1.88)       \\  \hline
    G2G &       46.25 ( 2.09)      &       46.61 ( 1.59)      &       46.61 ( 1.59)       &       10.36 ( 1.26)      &       14.55 ( 1.54)      &       14.55 ( 1.54)       \\ 
    TADW &       50.58 ( 2.37)      &       49.71 ( 1.99)      &       49.71 ( 1.99)       &       11.12 ( 0.56)      &       17.22 ( 2.18)      &       17.22 ( 2.18)       \\ 
    IDNE & 46.09 ( 0.98) & 46.51 ( 1.31) & 46.51 ( 1.31) & 9.51 ( 0.52) & 14.14 ( 0.31) & 14.14 ( 0.31) \\ \hline
    ATILP & \textbf{ 54.39 ( 1.76) } & \textbf{ 52.94 ( 0.84) } & 52.94 ( 0.84) & \textbf{ 14.11 ( 1.54) } & \textbf{ 22.02 ( 2.00) } & 22.02 ( 2.00) \\ 
\end{tabular}\end{adjustbox}

\vspace{1em}

\begin{adjustbox}{width=\columnwidth}
\begin{tabular}{ >{\centering\arraybackslash}p{2cm} | c  c  c | c  c  c } 
\textbf{ Science } & \multicolumn{3}{c|}{Inductive}  & \multicolumn{3}{c}{Transductive} \\
       &    \textbf{ AUC }    & \textbf{ P } &   \textbf{ R }   &    \textbf{ AUC }    & \textbf{ P } &   \textbf{ R }   \\ \hline
    random &       20.31 ( 1.28)      &       20.64 ( 1.08)      &       20.64 ( 1.08)       &        2.57 ( 0.34)      &        2.20 ( 0.83)      &        2.20 ( 0.83)       \\  \hline
    AT (title) &       30.07 ( 1.46)      &      \textbf{ 50.91 ( 1.57)}      &       31.40 ( 3.39)       &        4.65 ( 0.33)      &        9.85 ( 0.66)      &       29.55 ( 2.90)       \\ 
    AT (anchor) &       20.47 ( 1.50)      &       20.47 ( 1.50)      & \textbf{ 100.00 ( 0.00) } &        2.45 ( 0.19)      &        2.45 ( 0.19)      & \textbf{ 100.00 ( 0.00) } \\  \hline
    LSA &        50.82 ( 3.09)    &       49.22 ( 2.76)      &       49.23 ( 2.76)       &       11.70 ( 2.59)      &       17.59 ( 3.12)      &       17.61 ( 3.10)       \\  \hline
    DW &            -            &            -            &            -             &       11.11 ( 2.04)      &       16.38 ( 3.42)      &       16.38 ( 3.42)       \\  \hline
    G2G &       41.50 ( 3.50)      &       43.18 ( 3.64)      &       43.19 ( 3.66)       &        7.62 ( 0.95)      &       11.30 ( 1.31)      &       11.30 ( 1.31)       \\ 
    TADW &       48.65 ( 3.07)      &       48.32 ( 2.67)      &       48.33 ( 2.66)       &       10.70 ( 2.08)      &       \textbf{ 19.20 ( 3.10)}      &       19.23 ( 3.14)       \\ 
    IDNE & 41.11 ( 4.15) & 42.83 ( 3.12) & 42.83 ( 3.12) & 8.27 ( 1.47) & 11.06 ( 2.57) & 11.06 ( 2.57) \\ \hline
    ATILP & \textbf{ 51.01 ( 2.33) } & 50.36 ( 2.30) & 50.36 ( 2.30) & \textbf{ 13.07 ( 3.01) } & 18.56 ( 2.86) & 18.56 ( 2.86) \\ 
\end{tabular}\end{adjustbox}

\vspace{1em}

\begin{adjustbox}{width=\columnwidth}
\begin{tabular}{ >{\centering\arraybackslash}p{2cm} | c  c  c | c  c  c } 
\textbf{ The Little Prince }  & \multicolumn{3}{c|}{Inductive}  & \multicolumn{3}{c}{Transductive} \\
      &    \textbf{ AUC }    & \textbf{ P } &   \textbf{ R }   &    \textbf{ AUC }    & \textbf{ P } &   \textbf{ R }   \\ \hline
    random &       21.44 ( 0.87)      &       21.16 ( 0.70)      &       21.16 ( 0.70)       &        2.85 ( 0.07)      &        2.39 ( 0.47)      &        2.39 ( 0.47)       \\  \hline
    AT (title) &       33.39 ( 1.49)      &       44.58 ( 1.75)      &       51.01 ( 2.24)       &        5.70 ( 0.24)      &        8.35 ( 0.29)      &       53.37 ( 1.11)       \\ 
    AT (anchor) &       21.67 ( 1.17)      &       21.67 ( 1.17)      & \textbf{ 100.00 ( 0.00) } &        2.66 ( 0.05)      &        2.66 ( 0.05)      & \textbf{ 100.00 ( 0.00) } \\  \hline
    LSA &       53.84 ( 1.69)     & 51.58 ( 0.95) &       51.60 ( 0.94)       &       16.25 ( 0.63)      & 24.84 ( 1.84) &       24.88 ( 1.83)       \\  \hline
    DW &            -            &            -            &            -             &       15.87 ( 0.66)      &       22.70 ( 1.56)      &       22.75 ( 1.58)       \\  \hline
    G2G &       44.42 ( 1.23)      &       43.94 ( 1.53)      &       43.94 ( 1.53)       &        8.03 ( 0.38)      &       11.54 ( 0.45)      &       11.55 ( 0.47)       \\ 
    TADW &       52.05 ( 1.53)      &       49.98 ( 0.30)      &       50.00 ( 0.30)       &       15.08 ( 0.77)      &       23.59 ( 1.44)      &       23.62 ( 1.48)       \\ 
    IDNE & 46.78 ( 1.20) & 46.33 ( 1.85) & 46.33 ( 1.85) & 10.52 ( 0.95) & 16.82 ( 1.50) & 16.82 ( 1.50) \\ \hline
    ATILP & \textbf{ 54.55 ( 1.99) } & \textbf{ 52.01 ( 1.44) } & 52.01 ( 1.44) & \textbf{ 17.79 ( 0.14) } & \textbf{ 25.52 ( 1.57) } & 25.52 ( 1.57) \\ 
\end{tabular}\end{adjustbox}

\vspace{1em}

\begin{adjustbox}{width=\columnwidth}
\begin{tabular}{ >{\centering\arraybackslash}p{2cm} | c  c  c | c  c  c } 
\textbf{ Cristiano Ronaldo } & \multicolumn{3}{c|}{Inductive}  & \multicolumn{3}{c}{Transductive} \\
       &    \textbf{ AUC }    & \textbf{ P } &   \textbf{ R }   &    \textbf{ AUC }    & \textbf{ P } &   \textbf{ R }   \\ \hline
    random &       20.97 ( 2.17)      &       21.17 ( 2.23)      &       21.17 ( 2.23)       &        2.18 ( 0.17)      &        0.75 ( 0.39)      &        0.75 ( 0.39)       \\  \hline
    AT (title) &       33.44 ( 1.03)      &       45.34 ( 2.08)      &       51.20 ( 3.32)       &        5.38 ( 0.28)      &        7.75 ( 0.43)      &       57.40 ( 2.56)       \\ 
    AT (anchor) &       20.89 ( 2.23)      &       20.89 ( 2.23)      & \textbf{ 100.00 ( 0.00) } &        2.19 ( 0.18)      &        2.19 ( 0.18)      & \textbf{ 100.00 ( 0.00) } \\ \hline
    LSA & \textbf{ 48.17 ( 0.18) } &       48.00 ( 2.29)      &       48.00 ( 2.29)       &        9.92 ( 1.92)      &       16.05 ( 2.92)      &       16.11 ( 2.91)       \\ \hline
    DW &            -            &            -            &            -             &        9.35 ( 1.40)      &       14.68 ( 3.05)      &       14.68 ( 3.05)       \\  \hline
    G2G &       40.03 ( 1.86)      &       43.46 ( 0.65)      &       43.48 ( 0.67)       &        7.49 ( 0.63)      &       12.46 ( 1.71)      &       12.46 ( 1.71)       \\ 
    TADW &       45.98 ( 0.32)      &       46.37 ( 2.39)      &       46.38 ( 2.38)       &        9.04 ( 1.53)      &       14.99 ( 2.91)      &       15.02 ( 2.95)       \\ 
    IDNE & 37.97 ( 1.31) & 40.25 ( 0.98) & 40.25 ( 0.98) & 6.17 ( 0.83) & 10.12 ( 1.38) & 10.12 ( 1.38) \\ \hline
    ATILP & 48.16 ( 0.83) & \textbf{ 49.09 ( 1.68) } & \textbf{ 49.09 ( 1.68) } & \textbf{ 11.65 ( 2.10) } & \textbf{ 18.72 ( 3.09) } & 18.72 ( 3.09) \\ 
\end{tabular}\end{adjustbox}
\end{table}

\section{Conclusion and Future Work} \label{sec:co}

In this paper, we present the task of link prediction in the case of Wikipedia anchor text hyperlinks. We propose an evaluation procedure where the sampling method relies on a mapping from anchor texts to target documents. We evaluate several algorithms and we highlight that solving this problem requires modeling the interplay between the anchor texts, the source documents and the target documents. For future work, we want to use recent neural models in NLP to improve the modeling of these relations.  

\clearpage
\newpage

\begin{acks}
This work has been partially funded by the Agence Nationale pour la Recherche (ANR) through the following programs: ANR-19-CE38-0011 (ARCHIVAL), ANR-16-CONV-0002 (ILCB) and ANR-11-IDEX-0001-02 (A*MIDEX).
\end{acks}

\bibliographystyle{ACM-Reference-Format}
\balance
\bibliography{biblio}

\end{document}